\documentclass[twocolumn,showpacs,preprintnumbers,amsmath,amssymb]{revtex4}
\usepackage{graphicx}
\usepackage{dcolumn}
\usepackage{bm}
\begin{document}
\title{Phase space patterns of quantum transport on ordered and disordered networks}
\author{Xin-Ping Xu$^{1,2}$}
\author{Feng Liu$^{1}$}
\affiliation{ $^1$Institute of Particle Physics, HuaZhong Normal
University, Wuhan 430079, China \\
$^2$Institute of High Energy Physics, Chinese Academy of Science,
Beijing 100049, China }
\date{\today}
\begin{abstract}
In this paper, we consider the quantum-mechanical phase space
patterns on ordered and disordered networks. For ordered networks in
which each node is connected to its $2m$ nearest neighbors ($m$ on
either side), the phase space quasi-probability of Wigner function
shows various patterns. In the long time limit, on even-numbered
networks, we find an asymmetric quasi-probability between the node
and its opposite node. This asymmetry depends on the network
parameters and specific phase space positions. For disordered
networks in which each edge is rewired with probability $p>0$, the
phase space displays regional localization on the initial node.
\end{abstract}
\maketitle
\section{InTRoduction}
Quantum walk (QW), which is a generalization of the classical random
walk, has attracted a great deal of attention from the scientific
community. The continuous interest in the study of
quantum-mechanical transport process can be partly attributed to its
broad applications in the field of quantum information and
computation~\cite{rn1,rn2,rn3,rn4}. In recent years, two types of
quantum walks exist in the literature: the discrete-time quantum
coined walks and continuous-time quantum walks~\cite{rn5,rn6}. Both
discrete-time quantum coined walks and continuous-time quantum walks
have been argued to give an algorithmic speedup with respect to
their classical counterparts~\cite{rn7}.

In classical physics, the dynamical behavior of a system is
described by phase space variables, such as position and momentum. A
plot of position and momentum variables as a function of time shows
the phase diagram of classical transport process. In contrast to the
classical transport, the quantum-mechanical transport happens in the
Hilbert space~\cite{rn8}. Such transport process needs to be
formulated in phase space as a unified picture of the classical
transport. This can be done by the widely used Wigner function,
which transforms the wave function of a quantum mechanical state
into a function in the position-momentum space analogously defined
in the classical phase space~\cite{rn9}. It is shown in
Ref.~\cite{rn10} that integrating the Wigner function along the
lines in phase space is a positive value of probability and gives
the correct marginal distributions. However, the negativity of
Wigner function provides an indication of non-classical behavior.

The phase space method (Wigner function) provides a very useful tool
for the study of quantum states in the field of statistical
mechanics, quantum chemistry, molecular dynamics, scattering theory,
quantum optics, etc~\cite{rn11,rn12,rn13,rn15,rn16,rn17,rn18}. There
are various approaches available to generalize the Wigner function
for quantum systems with a finite-dimensional space of
states~\cite{rn19}. In Ref.~\cite{rn9}, Wootters have introduced a
discrete version of the Wigner function that has all the desired
properties only when $N$ is a prime number. The phase space defined
by Wootters is an $N\times N$ grid ($N$ is prime) and a Cartesian
product of such spaces corresponding to prime factors of $N$ in the
most general case~\cite{quant-com}. Notably, a more general discrete
Wigner function defined for a system with arbitrary values of $N$
has been introduced by Hannay and Berry in the studies of
semiclassical properties of classically chaotic systems~\cite{rn20}.
This version of Wigner function is used in several contexts and
recently applied to analyze the phase-space representation of
quantum computers and algorithms~\cite{quant-com,qc-pla,qt-pra}. In
this case, the phase space is constructed as a grid of $2N\times 2N$
points where the state is represented in a redundant manner (only
$N\times N$ of them are independent)~\cite{quant-com}. Recently,
M\"{u}lken \emph{et al} propose a version of discrete phase space of
Wigner function, which is defined for continuous-time quantum walks
(CTQWs) on a one-dimensional discrete network with periodic boundary
conditions~\cite{rn10}. This kind of discrete Wigner function
recovers the correct marginal distributions when it is added over
the horizonal and vertical lines. However, it is not positive when
added over the general lines~\cite{quant-com} in phase space. This
is an unique feature differs from the usual Wigner function, which
is positive when added over any lines~\cite{qt-pra}.

Here, we use the version of discrete phase space of Wigner function
proposed by M\"{u}lken~\cite{rn10}. In Ref.~\cite{rn10}, M\"{u}lken
\emph{et al} formulate CTQWs in phase space on a network of size $N$
whose nodes are enumerated as $0,1,...,N-1$. The Wigner function has
the form of a Fourier transform as follows~\cite{rn10},
\begin{equation}\label{eq1}
W(x,k,t)=\frac{1}{N}\sum_{y=0}^{N-1}e^{\frac{2i\pi ky}{N}}<x-y|
\hat{\rho}(t)|x+y>,
\end{equation}
Where $\hat{\rho}(t)$ is the density operator of a pure state, $k$
and $x$ denote the phase space coordinate of positions. The
summation over $y$ in the interval [$0,N-1$) can be carried out in
any $N$ consecutive values. Considering an initial exciton begins at
node $j$, the time evolution of the associated state $|j>$ is given
by $|j,t>=e^{-iHt}|j>$. Suppose $E_n$ and $|q_n>$ are the $n$th
eigenvalue and  eigenstate of the Hamiltonian of Laplace matrix, the
time independent Schr\"{o}dinger equation is $H|q_n>=E_n|q_n>$,
where $|q_n>$ spans the whole accessible Hilbert space and forms an
orthonormal complete basis set, i.e., $<q_n|q_l>=\delta_{nl}$,
$\sum_n|q_n><q_n|=1$. Inserting the complete set condition of the
eigenstates into the time evolution equation, we get,
\begin{equation}\label{eq2}
|j,t>=e^{-iHt}|j>=\sum_n e^{-iE_nt}|q_n><q_n|j>.
\end{equation}
The density operator of the system is $\hat{\rho}(t)=|j,t><t,j|$.
Substituting the above Equation into the Wigner function, we have,
\begin{equation}\label{eq3}
\begin{array}{ll}
&W_j(x,k,t)=\frac{1}{N}\sum_{y=0}^{N-1}e^{\frac{2i\pi
ky}{N}}\sum_{n,l}e^{-it(E_n-E_l)}\\
&\cdot <x-y|q_n><q_n|j><j|q_l><q_l|x+y>.
\end{array}
\end{equation}

In this paper, we use the above Equation to consider phase space
patterns of CTQWs on ordered and disordered networks. For ordered
networks, the topology organizes in a very regular manner, i.e.,
each node of the network is connected to its $2m$ nearest neighbors
($m$ on either side). For disordered networks, we employ the famous
WS model~\cite{rn21}, which triggers a surge of research of
small-world networks in the field of complex
networks~\cite{rn22,rn23,rn24,rn25,rn26}. In the WS
model~\cite{rn21}, each connection of the regular networks is
rewired with probability $p$. Tuning the rewiring probability $p$
interpolates the network topology between order ($p=0$) and disorder
(or random with $p=1$). The intermediate value $0<p<1$ corresponds
to the small-world region~\cite{rn21}.

The paper is organized as follows. In
Sec.~\uppercase\expandafter{\romannumeral 2}, we consider the phase
space patterns on ordered networks which correspond to the WS model
with rewiring probability $p=0$. In
Sec.~\uppercase\expandafter{\romannumeral 3}, we consider the phase
space patterns on disordered networks. Conclusions and discussions
are given in Sec.~\uppercase\expandafter{\romannumeral 4}.
\section{Phase space patterns on ordered networks}
In this section, we consider the quantum-mechanical phase space on
one-dimension regular networks of $N$ nodes in which each node is
connected to its $2m$ nearest neighbors ($m$ on either side). This
generalized regular network has broad applications in various
coupled dynamical systems, including biological
oscillators~\cite{rn27}, Josephson junction arrays~\cite{rn28},
neural networks~\cite{rn29}, synchronization~\cite{rn30},
small-world networks~\cite{rn31} and many other self-organizing
systems. We compute the phase space distribution on such general
network with periodic boundary conditions in the framework of Bloch
ansatz~\cite{rn32}, which is commonly used in solid state physics.
\subsection{Bloch ansatz and Wigner function}
The Hamiltonian (H) of the system for CTQWs is related to the
Laplace matrix (A) of the connected networks as $H=\gamma A$. Here,
for the sake of simplicity, we assume the transmission rate $\gamma$
for all the connections to be equal.  The nondiagonal elements
$A_{ij}$ equal to $-1$ if nodes $i$ and $j$ are connected and $0$
otherwise. The diagonal elements $A_{ii}$ equal to the number of
total links connected to node $i$, i.e., $A_{ii}$ equals to the
degree of node $i$. Therefore, the Laplace matrix $A$ of ordered
networks takes the following form,
\begin{equation}\label{eq4}
 A_{ij}=\left\{
\begin{array}{ll}
2m,   & {\rm if} \ i=j,\\
-1,   & {\rm if} \ i=j\pm z, z\in [1,m] \\
0,    & Otherwise.
\end{array}
\right.
\end{equation}
The Hamiltonian acting on the state $|j>$ can be written as
\begin{equation}\label{eq5}
H|j>=(2m+1)|j>-\sum_{z=-m}^m|j+z>, \ z \in Integers.
\end{equation}
The above Equation is the discrete version of the Hamiltonian for a
free particle moving on the network. Using the approach of Bloch
function~\cite{rn32} in solid state physics, the time independent
Schr\"{o}dinger equation reads
\begin{equation}\label{eq6}
H|\psi_n>=E_n|\psi_n>.
\end{equation}
The Bloch states $|\psi_n>$ can be expanded as a linear combination
of the states $|j>$ localized at node $j$~\cite{rn32},
\begin{equation}\label{eq7}
|\psi_n>=\frac{1}{\sqrt{N}}\sum_{j=0}^{N-1} e^{i\theta_n j}|j>.
\end{equation}
Substituting Eqs.~(\ref{eq5}) and (\ref{eq7}) into Eq.~(\ref{eq6}),
we obtain the eigenvalues (or energy) of the system,
\begin{equation}\label{eq8}
E_n=2m-2\sum_{j=1}^m \cos(j\theta_n).
\end{equation}
The periodic boundary condition for the network requires that the
projection of $|\psi_n>$ on the state $|N>$ equals to that on the
state $|0>$, thus $\theta_n=2n\pi/N$ with $n$ integer and $n\in[0,
N)$. Replacing $|q_n>$ by the Bloch states $|\psi_n>$ in
Eq.~(\ref{eq3}), we can get the Wigner function as follows,
\begin{equation}\label{eq9}
\begin{array}{ll}
&W_j(x,k,t)=\frac{1}{N^3}\sum_{y=0}^{N-1}e^{\frac{2i\pi
ky}{N}}\sum_{n,l}e^{-it(E_n-E_l)}\\
&\ \ \ \ \ \ \ \ \ \ \ \ \ \ \ \ \ \ \  \cdot e^{i\theta_n(x-y-j)}e^{i\theta_l(j-x-y)}\\
&=\frac{1}{N^3}\sum_{n,l}e^{-it(E_n-E_l)+i(\theta_n-\theta_l)(x-j))}\sum_{y=0}^{N-1}e^{\frac{2i\pi
(k-n-l)y}{N}}.
\end{array}
\end{equation}
The summation over $y$ can be written as $N\delta_{(N+k-n-l)}$,
where $\delta_{(N+k-n-l)}$ takes value $1$ if $(N+k-n-l)$ equals to
$0$ (or mod N) and $0$ otherwise. Thus the Wigner function can be
simplified as,
\begin{equation}\label{eq10}
W_j(x,k,t)=\frac{1}{N^2}\sum_{n=0}^{N-1}e^{-it(E_n-E_{N+k-n})}e^{\frac{2i\pi
(2n-k)(x-j)}{N}}.
\end{equation}
Substituting the expression of eigenvalues of Eq.~(\ref{eq8}) into
the above equation, we obtain discrete Wigner function of CTQWs on
one-dimension ordered networks.
\subsection{Time evolution of the Wigner function}
\begin{figure}
\scalebox{0.5}[0.5]{\includegraphics{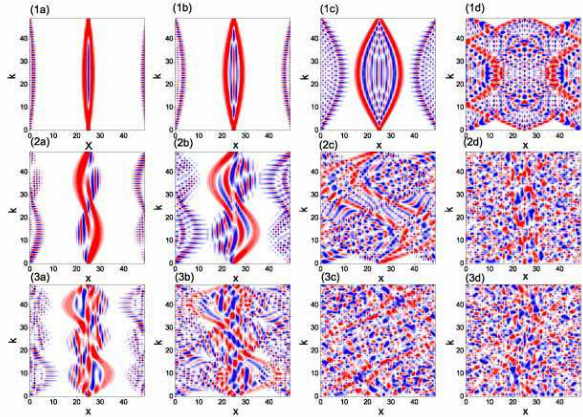}} \caption{(Color
online) Wigner functions $W_j(x,k,t)$ on networks of size $N=50$
with $m=1$ ((row 1), $m=2$ (row 2) and $m=3$ (row 3) at times
$t=1,2,5,20$ (columns (a)-(d)). The initial node is at $j=N/2=25$.
Red regions denote positive values of $W_j(x,k,t)$, blue regions
denote negative values and white regions close to value $0$.
\label{fg1}}
\end{figure}
We consider the time evolution of quantum-mechanical phase space
according to Eq.~(\ref{eq10}). Figure~\ref{fg1} shows a contour plot
of the Wigner function on ordered networks of size $N=50$ with
different values of $m$ at different times. At very small time
scales, the Wigner function is localized on the stripe at the
initial node and its opposite node. As time increases, the Wigner
function spreads over the whole network. On short time scales, the
Wigner function has a very regular structure until the wave fronts
between the initial node and opposite node start to interferes with
each other. Interestingly, the phase space structure is more complex
on highly connected networks compared to that on the cycle network
($m=1$), and the nodes are populated more quickly on highly
connected networks.

\begin{figure}
\scalebox{0.5}[0.5]{\includegraphics{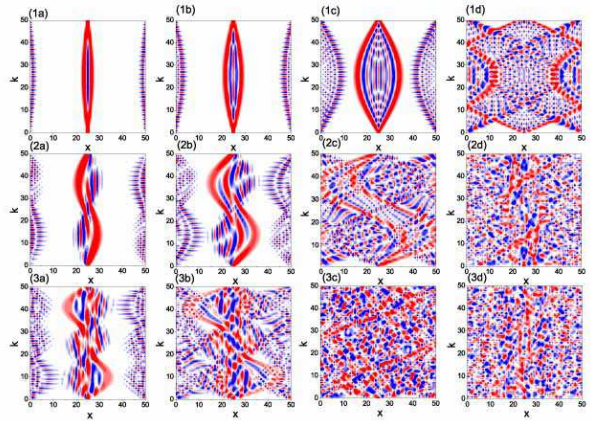}} \caption{(Color
online) Corresponding plots of Fig.~\ref{fg1} for $N=51$.
\label{fg2}}
\end{figure}

In Fig.~\ref{fg2}, we plot the Wigner function of CTQWs on networks
of size $N=51$. The phase space structure is quite similar to that
on even-numbered networks. We note that the phase space pictures
provide us more information of the underlying dynamics than the
transition probabilities. The phase space patterns alternate with
time frequently. At long time scales, the phase space becomes
irregular and we find that the structure of phase space has more
regularities on even-numbered networks than that on odd-numbered
networks with the same value of $m$ and $t$. This indicates the
higher topological symmetry of even-numbered networks.

It is worth noting that the pattern of Wigner function is
symmetrical about the axis for $m=1$. As $m$ increases, such
behavior disappears and $W_j(x,k,t)$ displays central symmetry at
the phase space center $x=k=j$. At the initial time $t=0$, on
even-numbered networks (or odd-numbered networks), the patterns of
Wigner function are the same for different values of $m$. This can
be concluded from Eq.~(\ref{eq10}). For the case of even $N$,
$W_j(j,k,0)$ equals to $1/N$ for arbitrary $k$. At the opposite node
$x=j+N/2$, $W_j(j+N/2,k,0)$ equals to $1/N$ for even $k$ and $-1/N$
for odd $k$. The nonzero values of the Wigner function at $x=j+N/2$
continue to show up at later times. The analysis for odd $N$ is
similar but the patterns are different. The Wigner function
$W_j(x,k,0)$ equals to $1/N$ for $x=j$ and $0$ otherwise. The
nonzero of $W_j(x,k,0)$ at the opposite node $x=j+N/2$ ($N\in$
Evens) is a natural consequence of the periodic boundary conditions
of the regular networks~\cite{rn10}. At later times, $W_j(x,k,0)$
involves contribution from all the eigenstates and the nodes
connected to the excitation node get populated, resulting the
semicircle-like areas on highly connected networks. We remark that
during all the time the patterns show central symmetry at $x=k=j$,
to this end, we conjecture that central symmetry is an intrinsical
feature of the phase space patterns.

\subsection{Long time averages}
The Wigner function of a specific position ($x,k$) fluctuates around
a constant value, thus it is interesting to study the long time
averaged phase space patterns. The time limiting Wigner function is
defined as,
\begin{equation}\label{eq11}
{\cal W}_j(x,k)=\lim_{T\rightarrow \infty}\frac{1}{T}\int_0^T
W_j(x,k,t)dt.
\end{equation}
For the ordered networks, the limiting Wigner function can be
simplified as,
\begin{equation}\label{eq12}
{\cal
W}_j(x,k)=\frac{1}{N^2}\sum_{n=0}^{N-1}\delta_{E_n,E_{N+k-n}}e^{\frac{2i\pi
(2n-k)(x-j)}{N}}.
\end{equation}
This expression shortens the numerical time of computation
considerably compared to Eq.~(\ref{eq11}). Therefore, we consider
the long time averaged phase space structure according to
Eq.~(\ref{eq12}).

For the cycle network ($m=1$) in which each node is only connected
to its two nearest neighbors, the limiting phase space has a simple
structure. For even-numbered networks ($N\in Evens$), the limiting
phase space structure is
\begin{equation}\label{eq13}
 {\cal
W}_j^e(x,k)=\left\{
\begin{array}{ll}
2/N^2,   & k\neq 0, k \in Evens \ and \ arbitrary \ x,\\
1/N,     & k=0 \ and \ x=j,j\pm N/2, \\
0,       & elsewhere.
\end{array}
\right.
\end{equation}

If the network size $N$ is an odd number, we can also obtain the
limiting Wigner function according to Eq.~(\ref{eq12}), which is
summarized as,
\begin{equation}\label{eq14}
 {\cal
W}_j^o(x,k)=\left\{
\begin{array}{ll}
1/N^2,   & k\neq 0 \ and \ arbitrary \ x,\\
1/N,     & k=0 \ and \ x=j, \\
0,       & elsewhere.
\end{array}
\right.
\end{equation}
which confirms the results in Ref.~\cite{rn33}.

For other values of $m$, the limiting phase space distributions can
also be calculated in the same way, but such process is complicated
for large value of $m$ because of the nontrivial degeneracy
distribution of the eigenvalues. Here, we report the numerical
results of phase space patterns on highly connected networks
according to Eq.~(\ref{eq12}).
\begin{figure}
\scalebox{0.9}[0.9]{\includegraphics{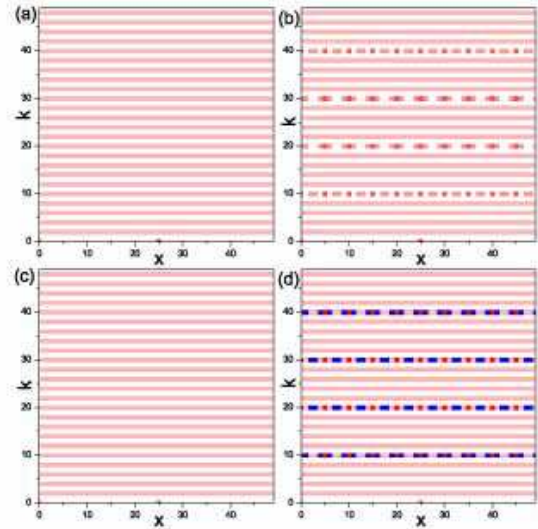}}
 \caption{(Coloronline)
Long time limiting Wigner function ${\cal W}_j(x,k)$ on ordered
networks of size $N=50$ for $m=1$ (a), $m=2$ (b), $m=3$ (c) and
$m=4$ (d). The initial node is at $j=N/2=25$. Red blocks denote
positive values of ${\cal W}_j(x,k)$, blue blocks negative values,
and the white regions denote zero.
 \label{fg3}}
\end{figure}

\begin{figure}
\scalebox{0.9}[0.9]{\includegraphics{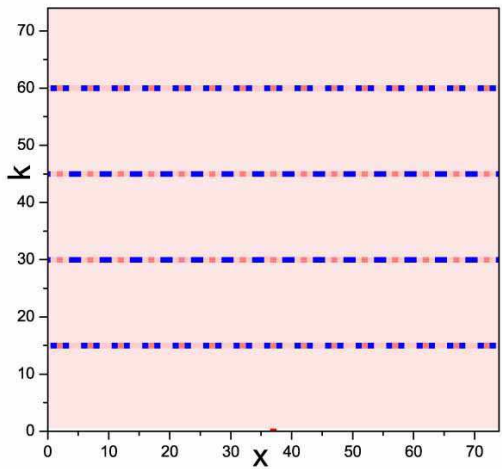}} \caption{(Coloronline)
Long time averaged phase space patterns on a network of $N=75$ and
$m=2$. The initial node is at $j=37$. Red blocks denote positive
values of ${\cal W}_j(x,k)$, blue blocks negative values, and the
white regions denote the value of $0$.
 \label{fg4}}
\end{figure}

Fig.~\ref{fg3} shows the phase space patterns on networks of $N=50$
with different values of $m$. We note that the phase space structure
of $m=1$ is the same as the phase space structure of $m=3$. After a
careful examination, we find that $m=1$ and $m=3$ have the same
degeneracy distributions of the eigenvalues. This explains the same
phase space patterns of $m=1$ and $m=3$. For $m=2$ and $m=4$, there
are some significant stripes in the phase space. Such nontrivial
stripes reflect the topological symmetry of the considered networks.
On the contrary, the phase space patterns on networks of $N=51$ with
$m=2$, $m=3$ and $m=4$ are the same as the structure with $m=1$,
which is described in Eq.~(\ref{eq14}). One may conjecture that the
phase space patterns do not alter when increasing the connectivity
$m$ on odd-numbered networks. But this is not true for some
particular value of network size $N$. For instance, on a network of
$N=75$ and $m=2$, we find significant stripes at some phase space
positions (See Fig.~\ref{fg4}).

The patterns of ${\cal W}_j(x,k)$ in Fig.~\ref{fg3} are the same for
odd $k$ and some even $k$. However, for some particular values of
$m$ and even $k$, the patterns are quite different (See the stripes
at $k=10,20,30$ and $40$ in Fig.~\ref{fg3}(b) and (d)). According to
Eq.~(\ref{eq12}), ${\cal W}_j(x,k)$ depends on the degeneracy of the
eigenstates, i.e., the Kronecker symbol $\delta_{E_n,E_{N+k-n}}$.
For $k=0$, $\delta_{E_n,E_{N+k-n}}$ equals to $1$ for all the values
of $n$, thus the sum in Eq.~(\ref{eq12}) contains $N$ exponential
terms, which results in ${\cal W}_j(x,0)=1/N$ for $x=j,j\pm N/2$ and
${\cal W}_j(x,0)=0$ otherwise. When $k$ is odd,
$\delta_{E_n,E_{N+k-n}}=0$ holds for all the values of $n$. For some
values of even $k$, $\delta_{E_n,E_{N+k-n}}$ equals to $1$ when
$n=k/2$ and $n=N/2+k/2$, this leads to ${\cal W}_j(x,k)=2/N^2$
($k\in$ arbitrary $x$). However, for some other values of even $k$,
$\delta_{E_n,E_{N+k-n}}$ does not vanish and ${\cal W}_j(x,k)$ is
oscillatory on the horizontal lines located at $k=10,20,30$ and
$40$. The oscillating strip is caused by the interference between
the positive strip and the mirror image. This is similar to the case
of decoherence of quantum walks in phase space~\cite{rn13}, in which
there are interference fringes in lines between the two position
eigenstates, and all the vertical lines have their corresponding
oscillatory counterparts originated from the boundary
conditions~\cite{rn13}.
\begin{figure}
\scalebox{0.5}[0.5]{\includegraphics{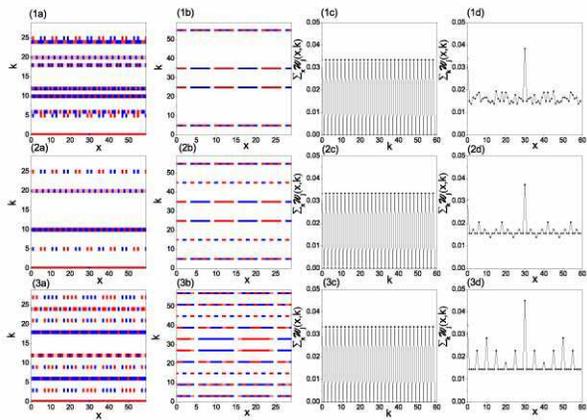}} \caption{
(a)Contour plot of ${\cal W}_j(x,k+N/2)-{\cal W}_j(x,k)$ on networks
of size $N=60$ with $m=2$ (row 1), $m=3$ (row 2) and $m=4$ (row 3).
(b): the same as column (a) but for ${\cal W}_j(x+N/2,k)-{\cal
W}_j(x,k)$ ($x\in [0,N/2)$). Red blocks denote positive values of
the discrepancy, blue blocks negative values, and the white regions
denote value of $0$ (symmetric quasi-probability in the phase
space). The last two columns show marginal distributions of
$\sum_x{\cal W}_j(x,k)$ (column (c)) and $\sum_k{\cal W}_j(x,k)$
(column (d)) for $m=2$, $m=3$ and $m=4$ (rows 1-3). The initial node
is at $j=N/2=30$.
 \label{fg5} }
\end{figure}

In Fig.~\ref{fg3}, we find a symmetric structure of the phase space
distributions in both the $x$ and $k$ directions. That is to say,
${\cal W}_j(x,k)$ equals to ${\cal W}_j(x+N/2,k)$ ($x\in [0,N/2)$)
for all the values of $k$; ${\cal W}_j(x,k)$ equals to ${\cal
W}_j(x,k+N/2)$ ($k\in [0,N/2)$) for all the values of $x$. Such
symmetric phase space structure exists on some even-numbered
networks. However, for some even-numbered networks with certain
values of $m$, this is not true. For instance, on networks of size
$N=60$ with $m=2$, $m=3$ and $m=4$, the Wigner function ${\cal
W}_j(x,k)$ differs from ${\cal W}_j(x+N/2,k)$ ($x\in [0,N/2)$) for
some values of $k$ and ${\cal W}_j(x,k)$ also differs from ${\cal
W}_j(x,k+N/2)$ ($k\in [0,N/2)$) for some values of $x$. Such
asymmetry depends on the specific phase space positions $(x,k)$. In
Ref.~\cite{rn33}, the authors find an asymmetry of transition
probabilities for the starting node and its mirror node on
two-dimensional networks, their definition of mirror node is based
on geometry symmetry of the network. In this paper, we define the
mirror node $x^{'}$ of a given node $x$ to be its opposite node,
i.e., $x^{'}=x+N/2$ (or $k^{'}=k+N/2$ in the longitudinal
direction). We study symmetric and asymmetric structure of the phase
space on the $x$ and $k$ direction separately. The asymmetric phase
space is particularly characterized by the difference between ${\cal
W}_j(x,k)$ and ${\cal W}_j(x+N/2,k)$ ($x\in [0,N/2)$) in the $x$
direction, and difference between ${\cal W}_j(x,k)$ and ${\cal
W}_j(x,k+N/2)$ ($k\in [0,N/2)$) in the $k$ direction. Therefore, we
use the quantities ${\cal W}_j(x+N/2,k)-{\cal W}_j(x,k)$ and ${\cal
W}_j(x,k+N/2)-{\cal W}_j(x,k)$ to detect the asymmetry of the phase
space. Columns (a) and (b) of Fig.~\ref{fg5} show these two
quantities in the phase space. The nonwhite blocks in the plot
indicate asymmetric phase space positions. From the figure we find
that the asymmetric structure in the phase space is very complex.
The phase space positions in which asymmetry occurs is dependant on
the precise values of network parameters $N$ and $m$, as well as the
specific phase space positions.

It is interesting to note that, in Figs.~\ref{fg1} and~\ref{fg3},
${\cal W}_j(x,k)$ has more identical values than $W_j(x,k,t)$ in the
phase space plane. $W_j(x,k,t)$ and ${\cal W}_j(x,k)$ display
central symmetry and mirror symmetry respectively. This suggests
that the limiting Wigner function ${\cal W}_j(x,k)$ has higher
symmetry in the structure compared to the instantaneous Wigner
function $W_j(x,k,t)$.
\begin{figure}
\scalebox{0.9}[0.9]{\includegraphics{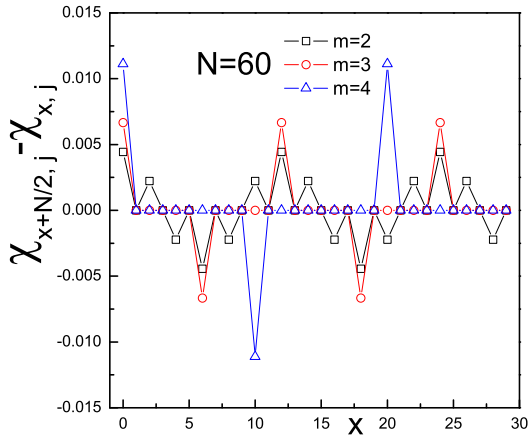}}
 \caption{(Color
online)Quantity $\chi_{x+N/2,j}-\chi_{x,j}$ as a function of $x$ on
networks of $N=60$ with different values of $m$. The initial node is
at $j=N/2=30$. The nonzero value of ($\chi_{x+N/2,j}-\chi_{x,j}$)
presents asymmetric probabilities between $\chi_{x+N/2,j}$ and
$\chi_{x,j}$.
 \label{fg6}}
\end{figure}
\subsection{Marginal distributions}
Summing the Wigner function along lines in phase space gives the
marginal distributions. Of course, such a process loses the detailed
information of the phase space. Here, we consider marginal
distributions of the long time averaged Wigner function summing over
the $x$ direction and $k$ direction respectively. Summing over all
$k$ recovers the time limiting transition probabilities, i.e.,
\begin{equation}\label{eq15}
\chi_{x,j}=\sum_k{\cal W}_j(x,k).
\end{equation}
Columns (c) and (d) of Fig.~\ref{fg5} show the marginal
distributions of the sum $\sum_x{\cal W}_j(x,k)$ and $\sum_k{\cal
W}_j(x,k)$. We find that the marginal distributions obtained by
summing over $x$ for $m=2$, $m=3$ and $m=4$ are the same (See column
(c) in Fig.~\ref{fg5}). $\sum_x{\cal W}_j(x,k)$ equals to $2/N$ for
even $k$ and $0$ for odd $k$. In contrast, the marginal
distributions $\chi_{x,j}$ for different values of $m$ are quite
different (See column (d) in Fig.~\ref{fg5}). Interestingly, there
is a large probability to be still or again at the initial node
($x=j$) and at the opposite node ($x=j-N/2$). In analogous to the
analysis of the phase space asymmetry, we consider asymmetry of the
limiting transition probability. We find that $\chi_{x+N/2,j}$
differs from $\chi_{x,j}$($x\in [0,N/2)$) for some particular values
of $x$ on networks with certain values of $N$ and $m$.
Fig.~\ref{fg6} shows the quantity $\chi_{x+N/2,j}-\chi_{x,j}$ for
all the values of $x$ in the range $[0,N/2)$. The nonzero values
$\chi_{x+N/2,j}-\chi_{x,j}$ in the plot indicate asymmetric
transition probabilities at the corresponding nodes.

As shown in Ref.~\cite{av-jpa}, the asymmetries of transition
probabilities originate from different contribution of eigenvalues
to $\chi_{x,j}$. It turns out that there are more contribution to
$\chi_{x,j}$ in the asymmetric cases than in the symmetric
cases~\cite{av-jpa}. The argumentation for our case is similar.
Although all the eigenvalues can be analytically obtained, a
complete analysis of all possible differences of eigenvalues
requires extensive work and is clearly beyond the scope of this
paper.
\section{phase space patterns on disordered networks}
In this section, we study the phase space patterns on networks in
the presence of two kinds of disorder: static disorder and
topological disorder. We import static disorder by adding a
perturbed Hamiltonian $\Delta$ to the original Hamiltonian $H$ as
done in Ref.~\cite{rn34}. This assumption do not create new
connections or remove the existing connections, thus does not change
the topology of the ordered networks. On the contrary for networks
with topological disorder in which each connection is rewired with
probability $p>0$, the topology of the networks becomes irregular.

In this paper, we focus on the influence of disorder on the long
time averaged phase space structure. Since the analytical
expressions for ordered networks do not apply any more, we
numerically calculate the Wigner function of CTQWs using the
software Mathematica. In order to save the computational time of the
numerical integrals, we combine Eqs.~(\ref{eq3}) and (\ref{eq11}) as
follows,
\begin{equation}\label{eq16}
\begin{array}{ll}
&{\cal W}_j(x,k)=\frac{1}{N}\sum_{y=0}^{N-1}e^{\frac{2i\pi
ky}{N}}\sum_{n,l}\delta_{E_n,E_l}\\
& \cdot <x-y|q_n><q_n|j><j|q_l><q_l|x+y>.
\end{array}
\end{equation}
In the following, we use the above Equation to report the numerical
results of Wigner function on networks with disorder.
\begin{figure}
\scalebox{0.5}[0.5]{\includegraphics{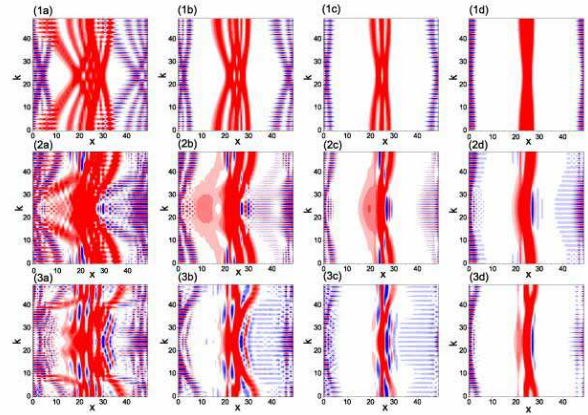}} \caption{(Color
online)Phase space patterns on networks of $N=50$ with exponential
disorder. The three rows are for $m=1$ (row 1), $m=2$ (row 2) and
$m=3$ (row 3) while the four columns correspond to the $\lambda
=0.5$ (column (a)), $\lambda =0.6$ (column (b)), $\lambda =0.7$
(column (c)) and $\lambda =0.8$ (column (d)). The initial node is at
$j=N/2=25$. Red regions denote positive values of ${\cal W}_j(x,k)$,
blue regions negative values and white regions denote values close
to $0$.
 \label{fg7}}
\end{figure}
\begin{figure}
\scalebox{0.5}[0.5]{\includegraphics{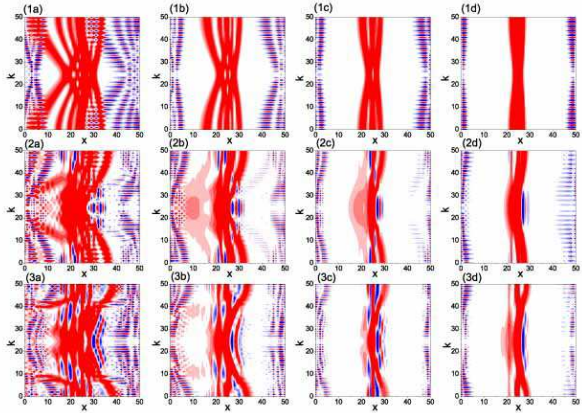}} \caption{(Color
online)The same plots as Fig.~\ref{fg7} but for $N=51$. \label{fg8}}
\end{figure}
\subsection{static disorder}
Analogous to the method in Ref.~\cite{rn34}, we consider the
exponential diagonal disorder whose off-diagonal elements of the
perturbed Hamilton $\Delta$ are $0$ and diagonal elements follow an
exponential function as,
\begin{equation}\label{eq17}
\Delta_{j,j}=e^{\lambda \frac{2\pi j}{N}},
\end{equation}
where $\lambda$ is the disorder parameter. The exponential disorder
in the above Equation has the advantage that the perturbed Hamilton
$\Delta$ possess a simple form in the Bloch representation, and
analytical solutions may be possible using the perturbation theory.
Here, we only give numerical results for the phase space structure
for such disorder. Figs.~\ref{fg7} and \ref{fg8} show the phase
space structure on networks of size $N=50$ and $N=51$ for different
values of $m$ and $\lambda$. The first column of the figures shows
${\cal W}_j(x,k)$ for $\lambda =0.5$ on networks with $m=1$, $m=2$
and $m=3$ (rows 1-3). For this weak disorder, the phase space
patterns have a very strange structure and the patterns of the
ordered networks are destroyed. Increasing the disorder parameter
$\lambda$, the patterns change drastically. The phase space
structure gets suppressed and a localized region forms at the
initial node $x=j$. In contrast to the patterns in Ref.~\cite{rn34},
the Wigner function here has negative values on highly connected
networks with large disorder (Compare the plots in the last column
to plots in Ref.~\cite{rn34}). The patterns in phase space on
odd-numbered networks (Fig.~\ref{fg8}) have a similar structure as
the even-numbered networks. However, differences are also visible
and it seems that there are more stripes on even-numbered networks
(Compare the corresponding plots in Fig.~\ref{fg7} and
Fig.~\ref{fg8}).

Compared to the patterns in Ref.~\cite{rn34}, the phase space
structure for networks with exponential disorder shows a strange
patterns. Such difference is induced by the distinct type of
disorder. The perturbed Hamiltonian in Ref.~\cite{rn34} has a Gauss
form, such disorder leads to a symmetrical phase space pattern in
the $x-k$ plane. Here, the exponential disorder has a heterogeneous
strength, i.e., sites labeled as large numbers have large
exponential disorder (See Eq.~(\ref{eq17})), this heterogeneous
disorder results in the complex pattern of the observed phase space.
For larger disorder, the difference of phase space between networks
with differen connectivity becomes invisible. We believe for
sufficient strong disorder, the phase space displays a complete
localization at the initial node.
\begin{figure}
\scalebox{0.5}[0.5]{\includegraphics{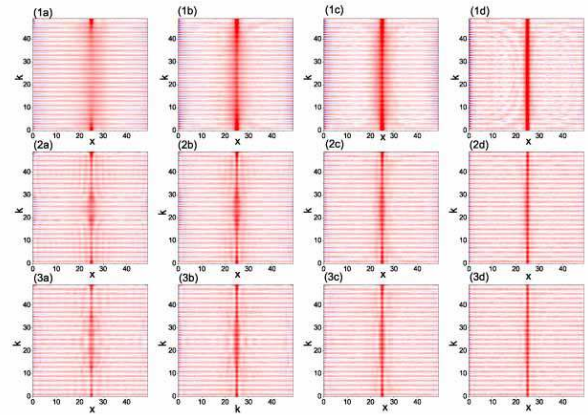}} \caption{(Color
online)Phase space patterns on WS networks of $N=50$ with $m=1$ (row
1), $m=2$ (row 2) and $m=3$ (row 3). The four columns correspond to
different values of disorder parameter $p$: $p=0.05$ (column (a)),
$p=0.1$ (column (b)), $p=0.2$ (column (c)) and $p=0.5$ (column (d)).
The initial node is at $j=N/2=25$ and all the plots are averaged
over $200$ realizations. Red regions denote positive values of
${\cal W}_j(x,k)$, blue regions negative values and white regions
denote values close to $0$. The colormaps are the same for all the
plots.
 \label{fg9}}
\end{figure}
\subsection{topological disorder}
\begin{figure}
\scalebox{0.5}[0.5]{\includegraphics{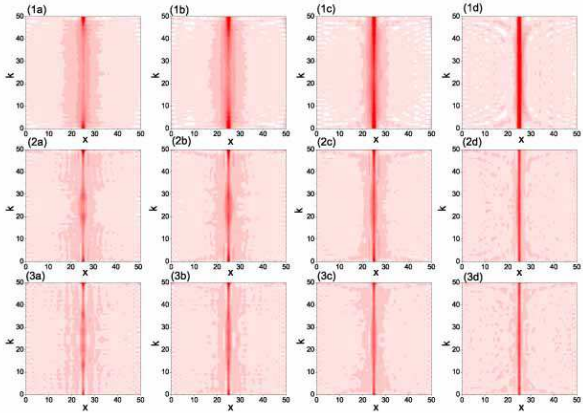}} \caption{(Color
online)The same plots as Fig.~\ref{fg9} but for $N=51$.
\label{fg10}}
\end{figure}

Since the topological structure of the WS model is not single, we
average the time limiting Wigner function over distinct
realizations. The ensemble average of the Wigner function provides
us a holistic view on the phase space structure of disordered
networks. Figs.~\ref{fg9} and~\ref{fg10} show the ensemble averaged
phase space patterns on disorder networks of size $N=50$ and $N=51$
with different values of $m$ and $p$. It is found that the stripes
on regular networks are destroyed and localized regions form around
the initial node $x=j$ (Compare Fig.~\ref{fg3} and Fig.~\ref{fg9}).
Increasing the rewiring probability $p$ (disorder parameter in the
WS model), the patterns change profoundly. The phase space structure
gathers together at the initial node $x=j$ for all $k$ on networks
with large disorder. However, the stripes on ordered networks are
still visible even for large disorder. When the disorder parameter
$p$ is fixed, the effect of localization becomes weak on networks
with more connectivity (Compare the plots in the same columns). For
odd-numbered networks, the patterns of localized central regions are
rather comparable to that on even-numbered networks. The other
regions in the phase space are analogous to the patterns of the
corresponding ordered networks, although the local patterns around
the $x\thickapprox j$ are changed.

\begin{figure}
\scalebox{0.9}[0.9]{\includegraphics{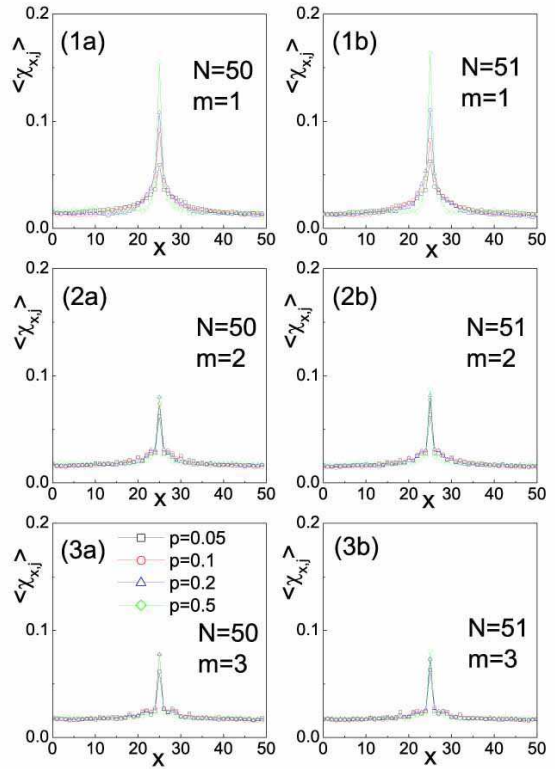}} \caption{(Color
online)Transition probability $<\chi_{x,j}>=\sum_k<{\cal W}_j(x,k)>$
on networks of size $N=50$ (column (a)) and $N=51$ (column (b)) with
$m=1$, $m=2$ and $m=3$ (rows 1-3). In each figure, the rewiring
probability $p$ takes values $0.05$ (squares), $0.1$ (circles),
$0.2$ (triangles) and $0.5$ (rhombus).
 \label{fg11}}
\end{figure}

Summing the averaged limiting Wigner function over $k$ gives the
averaged transition probability $<\chi_{x,j}>=\sum_k<{\cal
W}_j(x,k)>$. Fig.~\ref{fg11} shows this marginal distribution
$<\chi_{x,j}>$ for even and odd $N$ with different connectivity and
disorder strength. In the figure, we can see that there is a
remarkable localization at the initial node when $m=1$. As $m$
increases, the localizations become weak and nearly the same for
$m=2$ and $m=3$ (Compare the corresponding dots and curves in row 2
and 3). Interestingly, we note that the marginal distributions on
even and odd numbered networks are nearly the same and there is only
one peak at $x=j$. This feature differs from the case on ordered
networks where there are two peaks ($x=j$ and $x=j+N/2$) for even
$N$ and only one peak ($x=j$) for odd $N$. The two peak marginal
distribution on regular even-numbered networks is a consequence of
the symmetry of the network topology. When disorder was added, such
topological symmetry is destroyed, resulting in the single-peaked
marginal distributions on disordered networks.

\section{Conclusions and Discussions}
In conclusions, we have studied the quantum-mechanical phase space
patterns on ordered and disordered networks. On ordered networks
where each node connects to its $2m$ nearest neighbors ($m$ on
either side), the phase space quasi-probability of Wigner function
shows various patterns. In the long time limit, the phase space
structure presents different kinds of stripes. Such striped patterns
are related to the specific network size $N$ and connectivity
parameter $m$. Interestingly, if the network size $N$ is an even
number, we find an asymmetric quasi-probability and transition
probability between the node and its opposite node. This asymmetry
depends on the network parameters and specific phase space
positions. On disordered networks in which each edge is rewired with
probability $p>0$, the phase space displays regional localization on
the initial node.

The asymmetry of the quasi-probability and transition probability is
a novel phenomenon, which does not exist in the cycle graph with
$m=1$. However, we are unable to predict which particular parameters
of $N$ and $m$ or which phase space positions ($x,j$) are related to
such asymmetry. Is there relation between the phase space asymmetry
and transition probability asymmetry? Such question is interesting
and requires a further study. The phase space patterns on disordered
networks suggest that there are localizations in phase space on
small-world networks. Although both the static disorder and
topological disorder lead to localizations at the initial node,
localizations on small-world networks indicates that it is a generic
property of disordered systems and has important consequences for
quantum walk algorithms and quantum communication~\cite{rn35,rn36}.
\begin{acknowledgments}
The authors would like to thank Zhu Kai for converting the
mathematical package used in the calculations. This work is
supported by the Cai Xu Foundation for Research and Creation (CFRC),
National Natural Science Foundation of China under project 10575042
and MOE of China under contract number IRT0624 (CCNU).
\end{acknowledgments}

\end{document}